\documentclass[journal=nalefd,manuscript=letter,layout=twocolumn]{achemso}
\usepackage[T1]{fontenc} 
\usepackage[utf8x]{inputenc} 
\usepackage{hyperref}
\usepackage[version=3]{mhchem} % Formula subscripts using \ce{}
\usepackage{amsmath}
\usepackage{siunitx}
\usepackage{graphicx}
\usepackage{wasysym}
\usepackage{doi}

\let\oldmaketitle\maketitle
\let\maketitle\relax

\author{Arian Kriesch}
\email{arian.kriesch@mpl.mpg.de}
\affiliation[FAU and MPL]{Institute of Optics, Information and Photonics, Erlangen Graduate School in Advanced Optical Technologies, Friedrich-Alexander-University Erlangen-Nuremberg (FAU) and 
Max Planck Institute for the Science of Light (MPL), 91058 Erlangen, Germany}
\alsoaffiliation[Caltech]{Kavli Nanoscience Institute, California Institute of Technology, Pasadena, California 91125, USA}
\author{Stanley P. Burgos}
\affiliation[Caltech]{Kavli Nanoscience Institute, California Institute of Technology, Pasadena, California 91125, USA}
\author{Daniel Ploss}
\affiliation[FAU and MPL]{Institute of Optics, Information and Photonics, Erlangen Graduate School in Advanced Optical Technologies, Friedrich-Alexander-University Erlangen-Nuremberg (FAU) and 
Max Planck Institute for the Science of Light (MPL), 91058 Erlangen, Germany}
\author{Hannes Pfeifer}
\affiliation[FAU and MPL]{Institute of Optics, Information and Photonics, Erlangen Graduate School in Advanced Optical Technologies, Friedrich-Alexander-University Erlangen-Nuremberg (FAU) and 
Max Planck Institute for the Science of Light (MPL), 91058 Erlangen, Germany}
\author{\\Harry A. Atwater}
\affiliation[Caltech]{Kavli Nanoscience Institute, California Institute of Technology, Pasadena, California 91125, USA}
\author{Ulf Peschel}
\affiliation[FAU and MPL]{Institute of Optics, Information and Photonics, Erlangen Graduate School in Advanced Optical Technologies, Friedrich-Alexander-University Erlangen-Nuremberg (FAU) and 
Max Planck Institute for the Science of Light (MPL), 91058 Erlangen, Germany}
\title[SPP nanocircuits]
{Functional plasmonic nano-circuits with low insertion and propagation losses}

\begin{document}
	
%%%%%%%%%%%%%%%%%%%%%%%%%%%%%%%%%%%%%%%%%%%%%%%%%%%%%%%%%%%%%%%%%%%%%
%% The manuscript does not need to include \maketitle, which is
%% executed automatically.  The document should begin with an
%% abstract, if appropriate.  If one is given and should not be, the
%% contents will be gobbled.
%%%%%%%%%%%%%%%%%%%%%%%%%%%%%%%%%%%%%%%%%%%%%%%%%%%%%%%%%%%%%%%%%%%%%

\twocolumn[
\begin{@twocolumnfalse}
\oldmaketitle

\begin{abstract}
We experimentally demonstrate plasmonic nano-circuits operating as sub-diffraction directional couplers optically excited with high efficiency from free-space using optical Yagi-Uda style antennas at $ \lambda_0 = \SI{1550}{nm}$. The optical Yagi-Uda style antennas are designed to feed channel plasmon waveguides with high efficiency ($45 \%$ in coupling, $60 \%$ total emission), narrow angular directivity ($< \ang{40}$) and low insertion loss. SPP channel waveguides exhibit propagation lengths as large as \SI{34}{\micro m} with adiabatically tuned confinement, and are integrated with ultra-compact ($5 \times \SI{10}{\micro m}^2$), highly dispersive directional couplers, which enable \SI{30}{dB} discrimination over $\Delta \lambda = \SI{200}{nm}$ with only \SI{0.3}{dB} device loss.

\end{abstract}

\end{@twocolumnfalse}
]%
	\let\thefootnote\relax\footnote{$^\star$ To whom correspondence should be addressed}%
	\let\thefootnote\relax\footnote{$^\dagger$ FAU and MPL}%
	\let\thefootnote\relax\footnote{$^\ddagger$ Caltech}%
	\vspace{-\baselineskip}

%\section{Introduction and circuit design} % (fold)
\label{sec:introduction_and_circuit_design}
Surface plasmon polariton (SPP) waveguides are uniquely advantaged by their high confinement, allowing for subwavelength integration. This is a requirement for integrating optics with a footprint size that is comparable with electronic circuits - thus enabling plasmonic-electronic hybrid integration, a path that has been repeatedly highlighted as a future key application of plasmonics\cite{Engheta2007,Miller2009,Zia2006,Gramotnev2010}. 

However, high confinement in plasmonics usually increases loss due to the larger field overlaps with the metal. The second major obstacle to deep subwavelength plasmonics is high insertion loss due to the limited modal field overlap of less-confined waveguide schemes like Si integrated photonics\cite{Delacour2010,Asghari2011} or optical fibers\cite{Gosciniak2010}, thus intrinsically limiting the performance of hybrid dielectric-plasmonic circuits\cite{Briggs2010}.

\begin{figure}
  \includegraphics{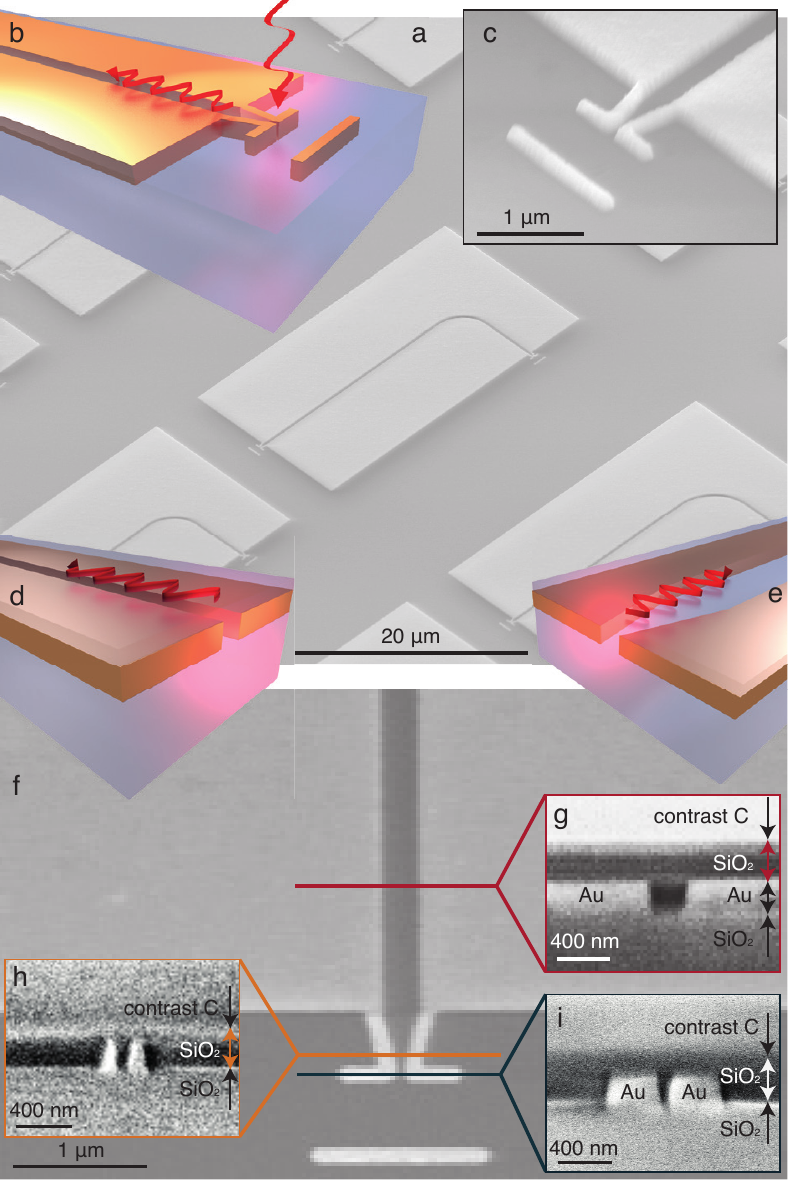}
  \caption{(a) SEM of an array of multicomponent plasmonic nano-circuits. (b, c) Optical loaded Yagi antennas are optimized to couple a highly focused, linearly polarized laser beam into waveguides (d) that are covered with a cladding layer of spin-on-glass. (e) Adiabatic tapers enable low-loss transitions between low-loss and high-confinement circuit sections. (f) The circuits are fabricated of Au with e-beam lithography. 
Cross-section cuts with a focused ion beam device (FIB), (g) through the \SI{300}{nm} wide waveguide, (h) the antenna connector and (i) the antenna with a minimum feature size of only \SI{80}{nm} display the sandwich of a substrate layer of silica, the \SI{220}{nm} thick metal structures and the \SI{320}{nm} thick cladding layer of spin-on-glass, which completely fills the metal gaps.}
  \label{fig:fig-1-sem}
\end{figure}

Here, we illustrate how in-circuit-loss can be mitigated by restricting strong optical confinement only to components where it is absolutely essential (Fig.~\ref{fig:fig-1-sem}a), and how insertion-loss can be addressed by coupling light into plasmonic nano-circuits via impedance-matched optimized Yagi-Uda\cite{Uda1927} style nano-antennas (Fig.~\ref{fig:fig-1-sem}b,c). Using this platform, we experimentally demonstrate optical directional couplers\cite{Yariv1973,Alferness1986} integrated on a micrometer scale that show unusually strong spectral dispersion, a key prerequisite for integrated wavelength division multiplexers.

To implement these device concepts, we use SPP channel waveguides\cite{Ly-Gagnon2012,Cai2010}, which offer maximum confinement\cite{Berini2011} in a narrow rectangular gap etched a few hundreds nanometers into a metal film. We note that this waveguide geometry does not suffer from optical mode cutoff when scaled down.

By filling the air gap ($n = 1$) in the plasmonic waveguide with the substrate material silica ($n \approx 1.45$) we make the modal field distribution more symmetric (Fig.~\ref{fig:fig-1-sem}d), therefore eliminating an upper modal cut-off that otherwise prohibits larger waveguide channel widths ($> \SI{80}{nm}$ for usual Au channel waveguides), therefore limiting the propagation length to below \SI{10}{\micro m}. 

Wherever possible, the connections from one functional plasmonic unit to the next must be bridged with low loss plasmonic waveguides. 
As the waveguide mode is basically maintained for different gap widths, easy to fabricate adiabatic waveguide tapers can form the transition from highly confining to low loss sections (Fig.~\ref{fig:fig-1-sem}e). 

We experimentally demonstrate that the investigated circuits achieve a propagation length of $L_{1/e} = \SI{34}{\micro m}$, while they are still subwavelength with a wave-guide gap width of \SI{300}{nm} and that they have an effective refractive index of $n_\text{eff} \approx 1.54$ at $\lambda_0 = \SI{1550}{nm}$ with low dispersion.

To reduce insertion loss, each presented nano-plasmonic circuit utilizes at least two connected Yagi-Uda antennas to enhance coupling efficiency from a focused laser beam and to achieve narrow directionality (Fig.~\ref{fig:fig-1-sem}c). We measured the antenna and waveguide properties spectrally and derive their fundamental properties: efficiency (\SI{45}{\%} in coupling, \SI{60}{\%} total emission), directionality ($< \ang{40}$) and spectral dispersion.

% section introduction_and_circuit_design (end)

%\section{Fabrication and measurement} % (fold)
\label{sec:fabrication_and_measurement}

For our experiments, embedded SPP nano-circuits (Figs. 1a and 6b) were fabricated in a two-step process. 
A \SI{220}{nm} thin Au film on a $160-\SI{}{\micro m}$-thin, polished quartz glass substrate was patterned by lift-off with a \SI{100}{kV} electron beam lithography system. The designed minimum feature size at the antenna gap (Fig.~\ref{fig:fig-1-sem}f) was \SI{80}{nm}.

Subsequently, flowable spin-on glass (Filmtronics 400F) was spin-coated and baked to form a dielectric cladding layer. 
Scanning electron microscope images of device cross-sections, cut with a focused ion beam (FIB) at different positions show that the cladding layer fills all gaps, covers the whole structure without cavities (Fig.~\ref{fig:fig-1-sem}g-i) and forms a smooth surface that is $\approx \SI{320}{nm}$ higher than the underlying structures. 

The fabricated circuits consistently deviate less than \ang{15} from perfectly rectangular sidewall angles. To investigate the dependence of the optical properties of each circuit component on design parameters, arrays of circuits were fabricated, while systematically varying waveguide length and bend radii (Fig.~\ref{fig:fig-1-sem}a).

Spectral transmission measurements of each circuit were taken on an optical far-field excitation and imaging setup\cite{Banzer2010a} with a \SI{4}{W} supercontinuum light source, spectrally filtered to  $\lambda = 1200 - \SI{1850}{nm}$ at \SI{5}{nm} FWHM with an acousto optical tunable filter (AOTF) (for sketch and details see supplementary information). 
A collimated and linearly polarized laser beam was focused by an objective ($NA = 0.9$ dry respectively $1.3$ oil immersion) to a diffraction-limited spot ($\diameter < \SI{2}{\micro m}$). The sample was positioned with \SI{5}{nm} resolution on a stabilized piezo stage (PI) to couple maximum power into the optical antennas (Fig.~\ref{fig:fig-1-sem}b). 

The whole circuit was simultaneously imaged with $150\times$ or $300\times$ magnification through the excitation objective, a polarization filter and NIR imaging optics on an infrared InGaAs CCD camera (Xenics XS, $320 \times 256$ pixels). 
With this setup, the linearly polarized emission from output antennas, turned by \ang{90} with respect to the input antenna (Fig.~\ref{fig:fig-1-sem}a), was imaged with enhanced dynamic range by suppressing the back-reflection from the incident beam with a ratio of up to $10,000$. 

Absolute circuit transmission values were determined by normalizing with the spectral reflectivity of a silica-air interface on the same sample (see supplementary information for details). 
With additional Fourier imaging, the polarization dependent angular emission pattern of the antennas was analyzed.

% section fabrication_and_measurement (end)

%\section{Yagi antennas} % (fold)
\label{sec:yagi_antennas}

\begin{figure}
  \includegraphics{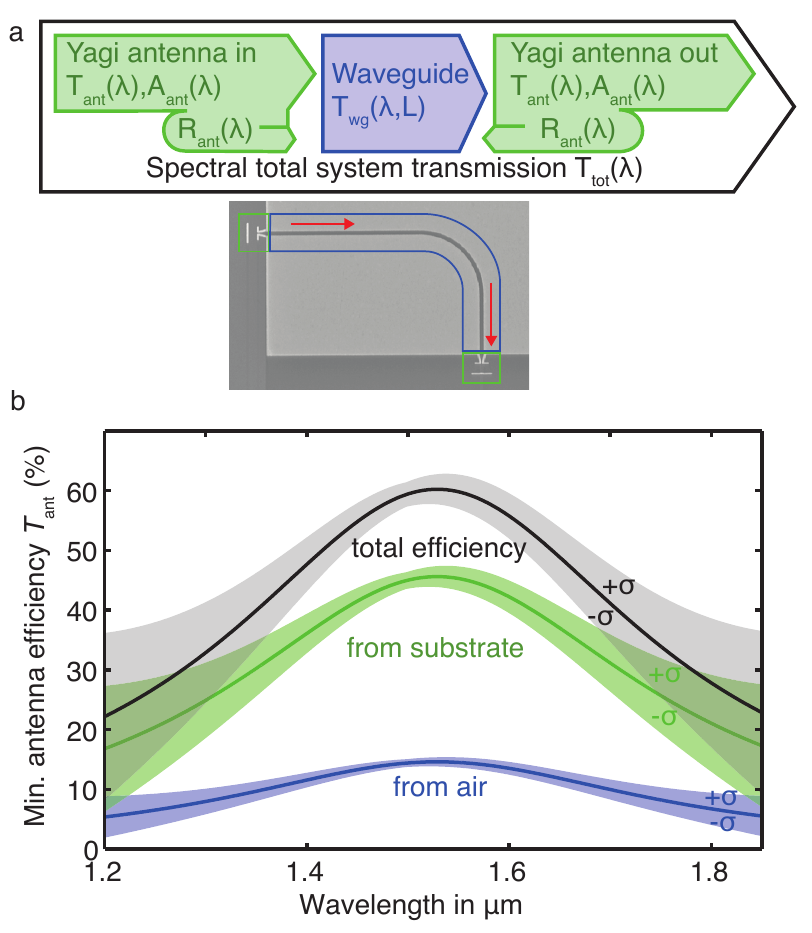}
  \caption{(a) The plasmonic circuits are analyzed as a sequence of black boxes, each with a characteristic, spectral transmission $T(\lambda)$, absorption loss $A(\lambda)$ and reflection coefficient $R(\lambda)$ towards the waveguide. 
(b) Experimentally determined lower bound for the spectral efficiency of the Yagi antennas $T_\text{ant}(\lambda)$ reaching a spectral peak value of \SI{60}{\%} of intensity to couple the bound mode inside the waveguide out into air (\SI{15}{\%}) and silica (\SI{45}{\%}), respectively from a focused beam into the waveguide. The antenna efficiency was determined by fitting measured data with a Lorentzian line shape and separating loss effects. 
The shaded areas indicate $\pm \sigma$ error bands.  }
  \label{fig:fig-2-antennaeff}
\end{figure}

\begin{figure*}
  \includegraphics{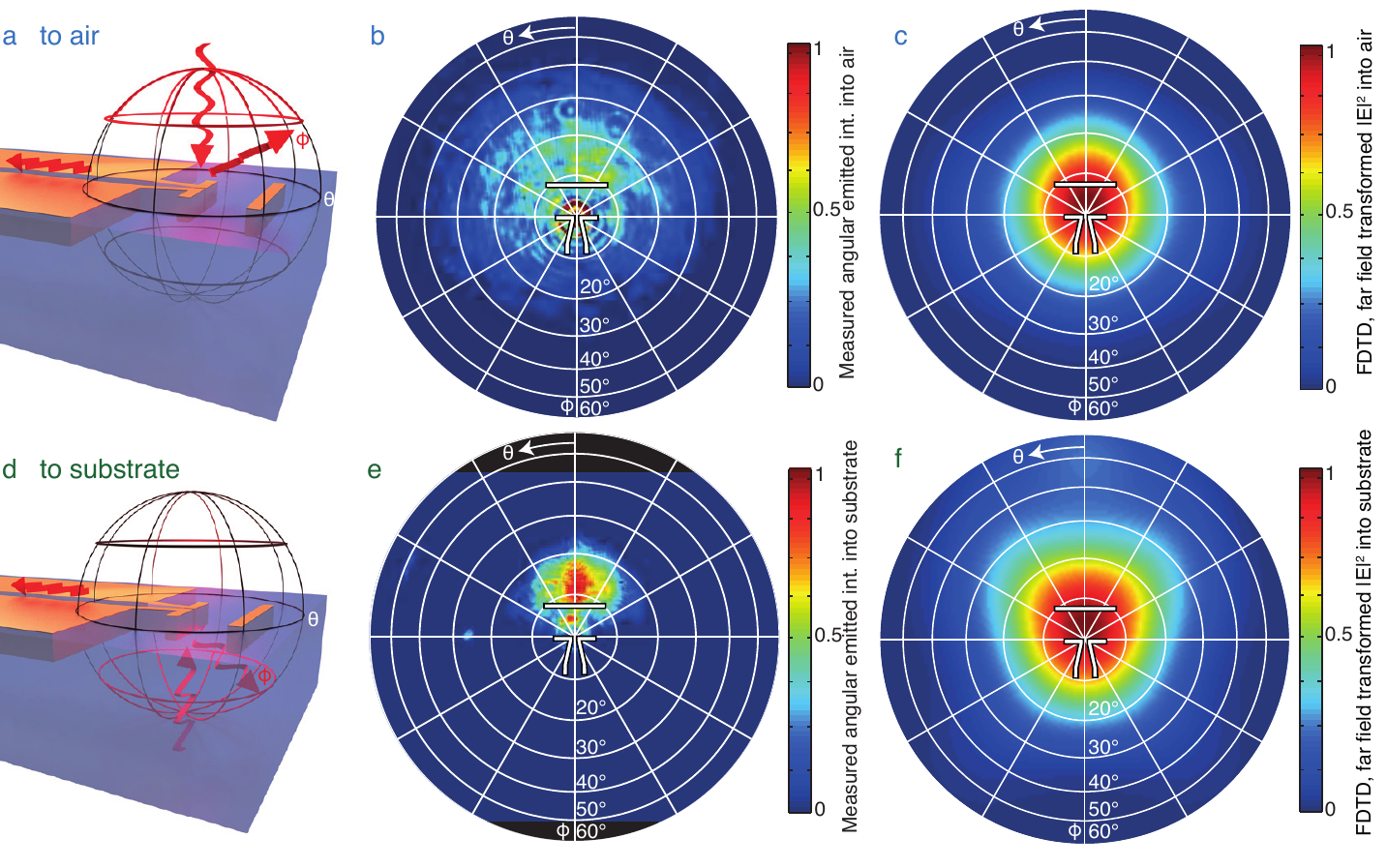}
  \caption{A narrow and strongly linear polarized angular emission directionality of the Yagi antenna at $\lambda = \SI{1550}{nm}$ (a,b,c) into air and (d,e,f) into the silica substrate was measured experimentally by Fourier plane imaging of the antenna emission during excitation with (b) an objective of $NA=0.9$ and (e) an immersion objective of $NA=1.3$.  The maximum azimuthal emission cone of \ang{30} is in accordance with the 3D FDTD simulation of the same structure into (c) air and (f) silica.}
  \label{fig:fig-3-angular}
\end{figure*}

For any application, including probing the circuit properties presented here, light has to be coupled into and out of the guided mode of the plasmonic waveguide with highest possible power conversion efficiency.\cite{Miller2009} Plasmonic antennas are best suited to fulfill this task for exciting channel plasmons from freespace.\cite{Novotny2011,Biagioni2012} 

The concept of optical antennas has already been successfully transferred from well-proven radiofrequency technology to the optical and near-infrared with a resulting antenna size in the \SI{}{\micro m} range.\cite{Wen2009a,Wen2011,Novotny2007} 
Here we apply the macroscopic design concept of Yagi-Uda antennas\cite{Uda1927} in the near infrared\cite{Andryieuski2012,Coenen2011,Curto2010,BernalArango2012,Dorfmuller2011} and connect them to feed the plasmonic circuits. 

For the investigated antennas, the feed element, an approximately \SI{1}{\micro m} long, \SI{100}{nm} wide, split Au rod, is centrally illuminated with a focused laser beam ($\diameter < \SI{2}{\micro m}$) that is polarized parallel to the antenna arms (Fig.~\ref{fig:fig-1-sem}b,c). 
This element is driven at its resonance frequency and feeds the \SI{300}{nm} wide SPP channel waveguide. An additional non-resonant Yagi-reflector element is placed in a distance of \SI{600}{nm} to the feed element to constructively reflect the electromagnetic field back into the waveguide, thus enhancing the in-coupling efficiency. 

Our whole design was developed using an iterative particle-swarm-optimization algorithm based on full 3D FDTD simulations while taking into account the limitations of fabrication and real optical material parameters as determined with ellipsometry. As it was found that in-coupling efficiency increases with decreasing size of the antenna gap21, we limited the antenna gap dimension to 80 nm based on fabrication limitations. A taper was added to connect the antenna with a 300-nm-wide low-loss waveguide to minimize impedance mismatch and back reflections. 

To simplify the experimental analysis of our circuits we attributed lumped properties to each element (Fig.~\ref{fig:fig-2-antennaeff}a)\cite{Engheta2007}. The antennas were assigned a characteristic spectral transmission $T_\text{ant}(\lambda) = P_\text{wg} (\lambda)/ P_\text{freespace} (\lambda)$ that represents the ratio of power, which is converted into the waveguide mode for a specific wavelength, a spectral absorption $A_\text{ant}(\lambda)$ due to Ohmic losses and a reflectivity $R_\text{ant}(\lambda)$ at the antenna-waveguide connection caused by residual impedance mismatch1,17. 

For the waveguides we assigned a length-dependent transmission $T_\text{wg}(\lambda,L) = 1 e^{-L/L_0}$ which is quantified in terms of the propagation length $L_0$ of their supported SPPs.
Hence, $L_0$ is separated by multiple measurements over a systematic variation of the waveguide length $L$. In applying this model to our experiments, we furthermore assume that in- and out-coupling antennas had similar, inversion-invariant properties.
This is a reasonable approximation as we inject and extract light using the same objective. Hence, the total system transmission is given by 

\begin{align*}
T_\text{tot}(\lambda) 
&= T_\text{ant1}(\lambda)  T_\text{wg}(\lambda)  T_\text{ant2}(\lambda) \\
&= T_\text{ant}(\lambda)^2  T_\text{wg}(\lambda).
\end{align*}

We experimentally measured the spectral system transmission $T_\text{tot}(\lambda)$ in the range of $\lambda = 1200 - \SI{1850}{nm}$ in steps of \SI{5}{nm}, coupling in and out of the circuit from air or through the silica substrate. An ensemble of 20 ($5 \times 5$) different plasmonic nano-circuits was characterized (Fig.~\ref{fig:fig-1-sem}a), repeated on 4 different samples with the same fabrication settings. Each parameter variation ensemble contained circuits with different total waveguide lengths ($L = 12.28 - \SI{42.28}{\micro m}$) and different radii ($R = 1 - \SI{4}{\micro m}$) for the 90° bend that turns the waveguide mode polarization for the emitting antenna. 

A systematic analysis of the bend loss over radius demonstrated that bends with $R = 3, \SI{4}{\micro m}$ offer negligible additional loss compared to the linear waveguide propagation loss, which is comprehensible since the modal effective wavelength difference for those radii is negligible compared to straight waveguides. Hence, we focused our analysis to circuits with a \SI{3}{\micro m} bend radius.

Based on this finding, we fitted the transmissions of the circuits with an iterative two-dimensional least-square fit over the independent variables, wavelength and waveguide length. The antenna transmission was modeled assuming a Lorentzian spectral broadening response, 
\begin{align*}
T_\text{ant}= \frac{ T_\text{ant}^\text{max}}
{  1+\left( (\lambda-\lambda_0)/\gamma \right)^2 }
.
\end{align*}
 
Under the assumption that dispersive effects of the in- and outcoupling are solely caused by the antenna resonance, we find that the peak wavelength $\lambda_0$ and linewidth $\gamma$ are independent of the excitation, whereas the maximum antenna efficiency $T_\text{ant}^\text{max}$ is higher for excitation from and emission into the substrate than into air due to the refractive index asymmetry of the structure, $n_\text{substrate} \approx 1.44 > n_\text{air} = 1$.

From this fit the spectral propagation length $L_0(\lambda)$ (Fig.~\ref{fig:fig-5-wg}a, blue curve) and the spectral antenna transmission $T_\text{ant}(\lambda)$ were determined over all measurements and circuits (few circuits with obvious fabrication errors were excluded from the evaluation). The fit error (standard deviation $\sigma$), therefore includes statistical errors of both the measurement and fabrication (Fig.~\ref{fig:fig-2-antennaeff}b). The determined antenna efficiency is a lower limit, as we might not have eliminated all possible sources of losses in the circuit. 

The antenna transmission peaks at $ \lambda_\text{max} = \SI{1530}{nm}$ with a spectral width of $\gamma = \SI{244}{nm}$, corresponding to a $FWHM = \SI{488}{nm}$. The spectral peak efficiency of the antennas is $T_\text{ant,air}^\text{max}=15\pm 1 \%$ from air and $T_\text{ant,substr}^\text{max} = 45 \pm 2 \% $ from the substrate with minimum statistical error around the peak and increasing error towards the measurement limits of $\lambda = \SI{1200}{nm}$ and $\lambda = \SI{1850}{nm}$, where the absolute circuit transmission and the InGaAs camera efficiency decrease.

Keeping the initial assumption that the antenna properties obey optical inversion symmetry, the total antenna efficiency, giving the total emission into air and silica, can be spectrally summated to be $T_\text{ant,tot}^\text{max} = 60 \pm 3 \%$, the highest power transmission for a waveguide loaded optical antenna, reported to date.

\begin{figure}
  \includegraphics{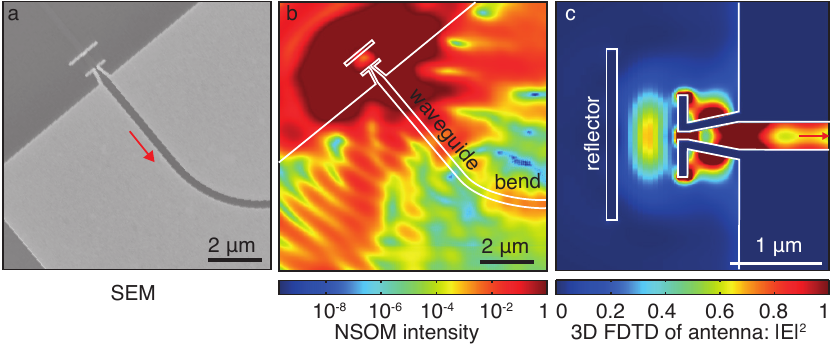}
  \caption{(a) SEM of a Yagi antenna that was illuminated with a highly focused beam through the substrate and (b) scanned with an aperture near field optical microscope (NSOM) through the \SI{300}{nm} thin layer of cladding $\text{SiO}_2$. Strong enhancement of the spatial near-field distribution is visible around the antenna and the Yagi reflector. The reflector suppresses emission of the antenna away from the waveguide and couples the electromagnetic wave into the SPP channel waveguide with only slight emission towards the cladding. (c) Electric field distribution ($|E|^2$) in the structure plane of the optimized antenna geometry from a full 3D FDTD simulation in the same configuration as the measurement.}
  \label{fig:fig-4-nsom}
\end{figure}

Validity of the applied transmission analysis is based on the hypothesis that the objective collects all light that is emitted from the Yagi antennas and that a good overlap of the focused exciting beam and the angular emission directionality is achieved. 

To verify this assumption, the antenna was excited from air (Fig.~\ref{fig:fig-3-angular}a) and from substrate (Fig.~\ref{fig:fig-3-angular}d), while the far-field emission of the antenna was imaged into the excitation direction (Figs.~\ref{fig:fig-3-angular}b,e) in the Fourier plane and compared to the angular emission spectrum obtained from 3D FDTD simulation with the same geometry parameters and excitation (Figs.~\ref{fig:fig-3-angular}c,f). 

In experiment and simulation the antenna emits with strong directionality within a polar angle cone of $\theta \leq \ang{40}$, which is remarkably narrow\cite{Biagioni2012,Curto2010} and completely covered by the NA of our objectives  ($NA = 0.9$ from air and $1.3$ immersion from silica). Hence, radiative losses seem to play a minor role in our measurements.

However, the guided field in a waveguide cannot be directly measured in the far-field, with the only limited information coming from slight scattering of impurities in strongly overexposed leakage microscopy (see supplementary). 
Therefore aperture near-field scanning optical microscopy (NSOM) was used to image the antennas and waveguides (Fig.~\ref{fig:fig-4-nsom}a) directly. 

The field distribution of the focused beam at the antenna, as predicted by 3D FDTD simulations (Fig.~\ref{fig:fig-4-nsom}c), leads to a distortion of the incident beam in the area between the antenna feed element and the reflector (Fig.~\ref{fig:fig-4-nsom}b). No propagating waves are visible along the surface of the sample away from the antenna, demonstrating the efficiency of the Yagi reflector. We note that the evanescent field of the guided mode inside the waveguide is still possible to image through the 320 nm thick cladding layer, which is an advantage for probing the operation of the nano-circuit dynamics.

% section yagi_antennas (end)

%\section{Waveguide properties} % (fold)

\begin{figure}
  \includegraphics{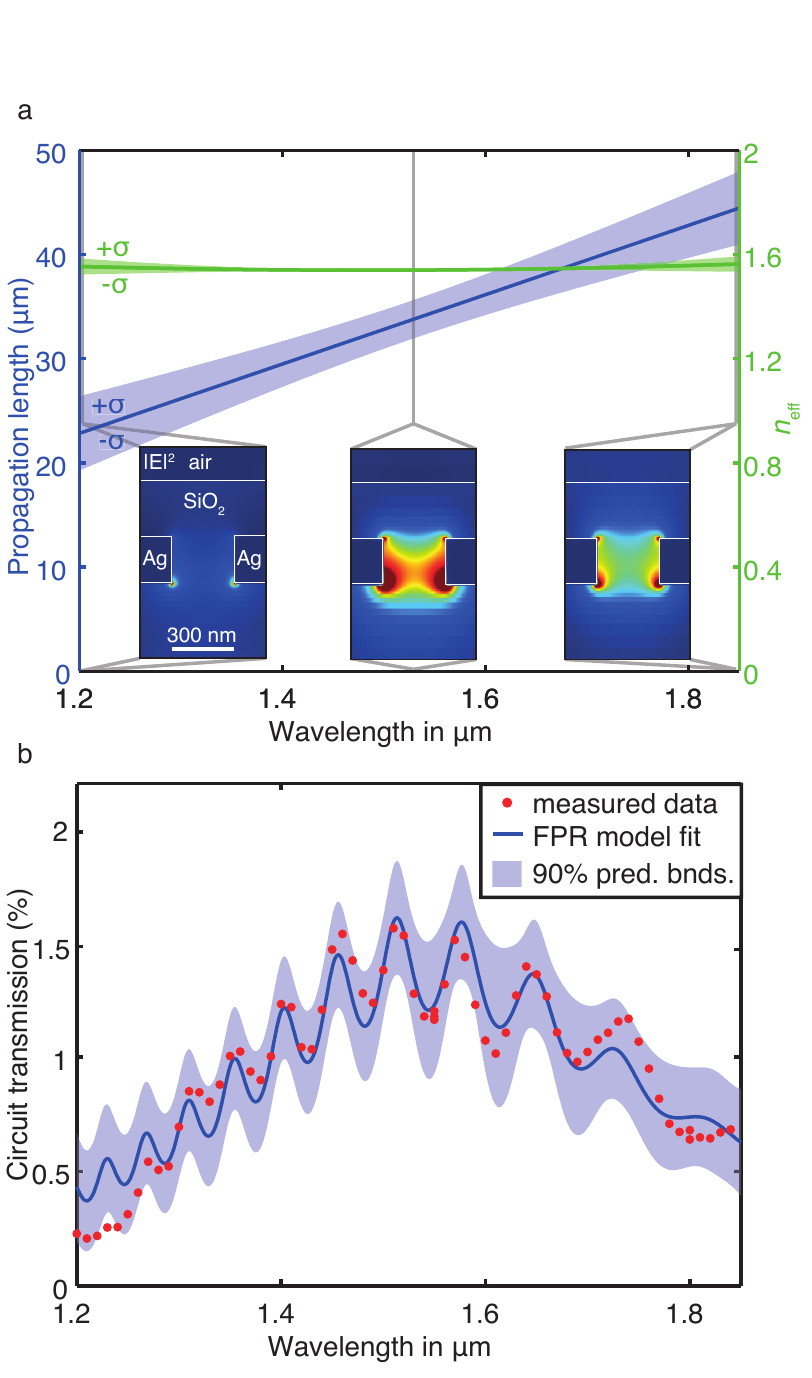}
  \caption{(a) The \SI{300}{nm} wide and \SI{220}{nm} deep embedded SPP channel waveguides offer remarkably low loss with a propagation length of  $P_0/e (\lambda = \SI{1550}{nm}) = \SI{34}{\micro m}$. Propagation length and dispersion (blue curve) were determined by iterative fitting to the experimental spectral circuit transmission for series of different waveguide lengths. The effective refractive index of the SPP mode was determined as $n_\text{eff} (\lambda = \SI{1550}{nm}) = 1.54$ with flat dispersion, by (b) iterative fitting of a Fabry-P\'erot oscillator model the experimental spectral transmission (see Fig.~\ref{fig:fig-3-angular}a). The shown pronounced oscillations (red dots) are caused by slight back-reflections of the wave inside the waveguide from the antennas in circuits with \SI{12.28}{\micro m} length.}
  \label{fig:fig-5-wg}
\end{figure}

\label{sec:waveguide_properties}
Embedded SPP channel waveguides feature a characteristic propagation length and effective refractive index. 
In particular the latter property is important for potential applications as it determines the phase velocity, but it is usually difficult to measure, as phase information is lost when measuring emitted intensities. 
From transmission measurements on circuits with short waveguides, we found spectral oscillations on the total transmission with maximum amplitude at the antenna resonance, which correspond to Fabry-P\'erot resonances, allowing us to probe the mode index of the waveguides.

For the shortest waveguides ($L = \SI{14.28}{\micro m}$), the spectral oscillations are very distinct (Fig.~\ref{fig:fig-5-wg}b). 
The lumped circuit transmission model (Fig.~\ref{fig:fig-2-antennaeff}a) allows us to treat the circuit as a Fabry-P\'erot resonator defined by a channel waveguide of length $L$ and mirrors formed by the antennas, due to imperfect antenna impedance-matching. 

This model was fitted to the spectral system transmission, taking the already determined waveguide loss (Fig.~\ref{fig:fig-5-wg}a,~blue curve) and the antenna efficiency into account. Hence, the effective refractive index of the waveguide mode (Fig.~\ref{fig:fig-5-wg}a,~green~curve) was determined to be $n_\text{eff}≈ 1.54$ at $\lambda_0 = \SI{1550}{nm}$ with low spectral dispersion, a value which coincides well with the expected guided mode effective index from FDTD calculations.
Simulated modal field distributions (Fig.~\ref{fig:fig-5-wg}a,~insets) indicate decreasing confinement and increasing field-overlap with the dielectric for longer wavelengths\cite{Dionne2006a}, thus explaining the experimentally observed decrease in propagation loss for these wavelengths (Fig.~\ref{fig:fig-5-wg}a,~blue curve).

% section waveguide_properties (end)

%\section{Optical directional couplers} % (fold)
\label{sec:optical_directional_couplers}

\begin{figure}
  \includegraphics{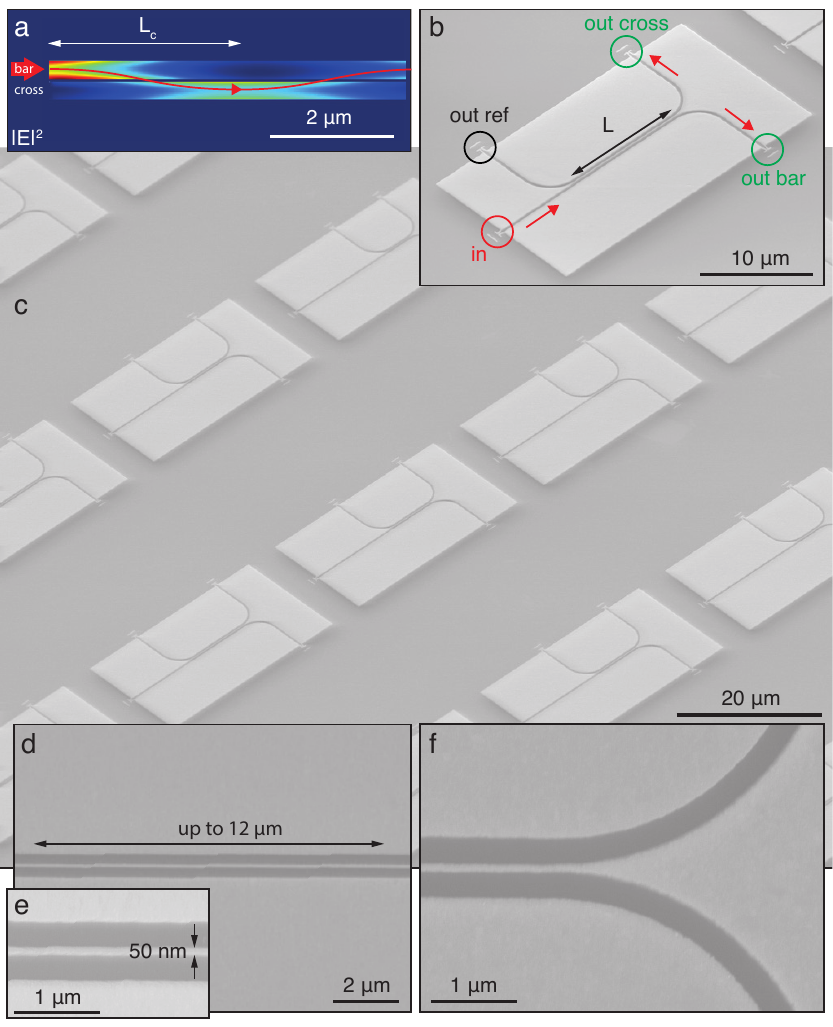}
  \caption{(a) Two embedded SPP channel waveguides in close vicinity exchange energy and form an optical directional coupler (ODC) with a subwavelength length $L_\text{c} = \SI{2.46}{\micro m}$ for full transfer of the displayed power flow from waveguide bar to cross (in-plane cross section of a 3D FDTD with the experimental geometry). 
  (b) For full spectral characterization of this passive SPP component, it was integrated into a nano-circuit and probed with 4 Yagi antennas, one to couple in (red), two emitting in crossed polarization to observe cross and bar (green) and one to monitor internal reflections (blue). 
  (c) The ODC parameters, length and waveguide separation were varied systematically. 
  (d, e) The thinnest separation is formed by an only $\SI{50}{nm}$ wide, $\SI{220}{nm}$ deep filament of Au, running uniformly over a length of up to $\SI{12}{\micro m}$. 
  (f) At the beginning and the end of the ODC the waveguides approach with a smooth, adiabatic bend that already causes significant coupling.}
  \label{fig:fig-6-sem-odc}
\end{figure}

We integrated the developed Yagi antennas and waveguides with optical directional couplers (ODC) in the developed nano-plasmonic circuit platform, demonstrating the functional application of sub-diffraction plasmonics as ultra-short coupling length devices.

ODCs have been a standard component in macroscopic integrated and fiber optics for several decades11. 
They allow for defined, dispersion-engineered transfer of power from one waveguide (bar) to a second waveguide (cross) by evanescent coupling of the field of the guided modes, which can be well described with coupled-mode-theory\cite{Yariv1973}. 
The coupling-ratio $I_\text{cross}/I_\text{bar}$ can be tuned with the length of the coupler. In the investigated nano-circuits, two embedded SPP waveguides run in parallel for up to \SI{12}{\micro m} (Fig.~\ref{fig:fig-6-sem-odc}b). 

3D FDTD simulations of the ODC (Fig.~\ref{fig:fig-6-sem-odc}a) clearly show that with thin metal filaments between the two waveguides the investigated geometry features coupling lengths $L_\text{c}$ for a first full power transfer from bar to cross down to few micrometers. 

Feed (Fig.~\ref{fig:fig-6-sem-odc}b,~red) and probe antennas (Fig.~\ref{fig:fig-6-sem-odc}b,~green) and an additional fourth antenna for monitoring internal back-reflections (Fig.~\ref{fig:fig-6-sem-odc}b,~blue) are connected to the different ports of the ODC. 
As before, the emission antennas are all turned by \ang{90}, radiating with a linear polarization perpendicular to the excitation antenna. The length of the couplers was varied in an array of optical circuits (Fig.~\ref{fig:fig-6-sem-odc}c) from $L = 0 - \SI{12}{\micro m}$ in \SI{3}{\micro m} steps in one direction, while the nominal width of the filament that separates the waveguides was varied from $w = 50 - \SI{90}{nm}$ in \SI{10}{nm} steps in the other. 

Even the smallest, \SI{50}{nm} thin metal filaments (aspect ratio $4.4$ for a \SI{220}{nm} metal film) demonstrated good fabrication fidelity over the entire \SI{12}{\micro m} length of the longest couplers (Figs.~\ref{fig:fig-6-sem-odc}d,e). 
To enable smooth transitions, the cross waveguide approaches the bar waveguide in a \ang{90} bend with a radius of $R = \SI{3}{\micro m}$, which for the single waveguide circuits showed negligible bend loss. Fabrication was consistently reproducible with 4 different samples. 

\begin{figure}
  \includegraphics{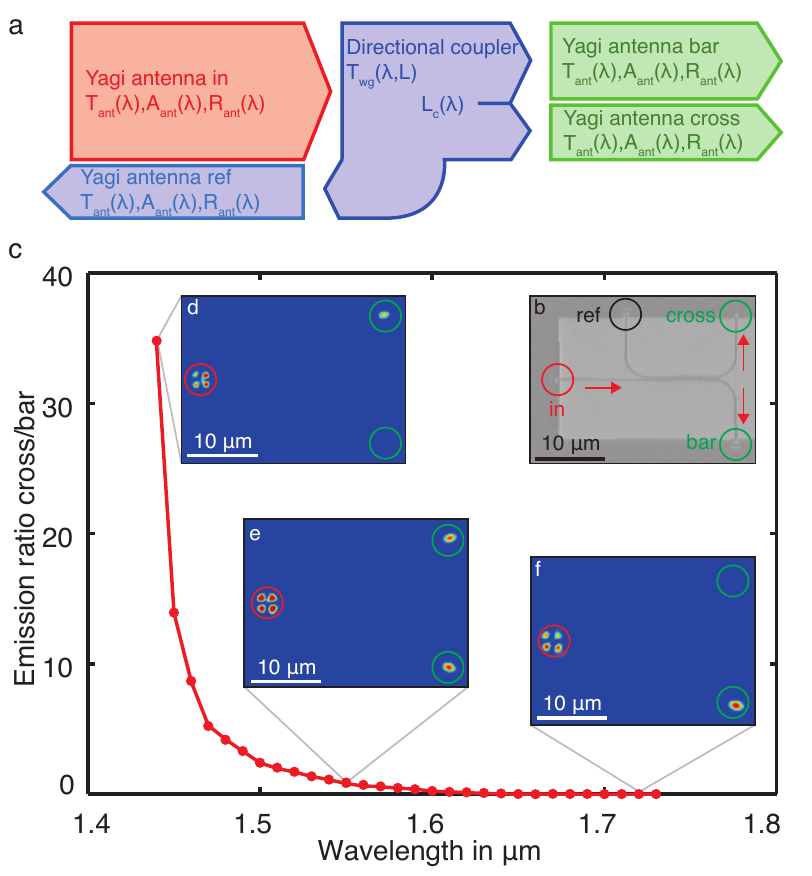}
  \caption{(a) Schematic diagram and (b) SEM of an ultra-compact SPP circuit with four Yagi antennas (red in, green out, black reflection reference) and a directional coupler (coupler length $L = \SI{10}{\micro m}$), designed for several full coupling cycles at $\lambda = \SI{1550}{nm}$. It shows strong spectral dispersion of the coupling length $L_\text{c}(\lambda)$. (c) This leads to full switching of the power from output channel cross to bar (green), clearly visible in the optical emission images for (d) \SI{1440}{nm}, (e) \SI{1550}{nm} and (f) \SI{1720}{nm}, resulting in a \SI{30}{dB} wavelength discrimination between \SI{1450}{nm} and \SI{1650}{nm}, while back-reflections (black) are negligible. }
  \label{fig:fig-7-odc}
\end{figure}

As for the basic straight waveguide system, the directional coupler was spectrally probed and modeled with a slightly more elaborate lumped circuit model (Fig.~\ref{fig:fig-7-odc}a) in which the previously investigated Yagi antennas are connected to the directional coupler with a characteristic spectral transmission $T_\text{wg}(\lambda,L)$ and coupling length $L_\text{c}(\lambda)$. Each circuit was probed at $\lambda_0 = \SI{1550}{nm}$, while two clearly distinct emission spots from the cross and bar antenna were separately integrate and the total system transmission was monitored similar as for the single waveguide circuits (Fig.~\ref{fig:fig-7-odc}e, see supplementary for details). 

We note that the resulting power ratio is intrinsically robust to variations of the coupling efficiency into the circuit as those variations simply lead to a linear scaling of the emission from both monitor antennas. 
Coupled mode theory predicts a power exchange between the waveguides in analogy to two weakly coupled damped harmonic oscillators\cite{Yariv1973}. 
An iterative least-square fit of this model was applied to the measured cross- and bar- emission, while being careful to avoid numerical divergences and ensuring equal weighting of all emission ratios (see supplementary information for details). 

Following this fit routine, we obtained a reproducible coupling length of only $L_\text{c} = 2.46 \pm \SI{0.04}{\micro m}$ for the thinnest filament width of \SI{50}{nm}. 
This fit, as a second free parameter, determines an additional equivalent length $L_\text{bend} = 1.7 \pm \SI{0.07}{\micro m}$ of the coupler that is caused by the transition into and out of the straight coupling region by \ang{90} bends (Fig.~\ref{fig:fig-6-sem-odc}). 

Hence, even ODCs with a parallel coupler length of $L = \SI{0}{\micro m}$ demonstrated significant coupling of the guided plasmon from the bar waveguide over to the cross waveguide. 
Furthermore, direct comparison of experimental and FDTD results demonstrate that the fabricated filament widths are effectively a bit smaller than expected, which is reasonable since the metal side-walls are inclined $\approx \ang{15}$ from the vertical, as is evident from SEM cross-sectional images of the fabricated device (Figs.~\ref{fig:fig-1-sem}g,h,i).

Consequently, ODCs simulated with a rectangular filament of \SI{40}{nm} width (Fig.~\ref{fig:fig-6-sem-odc}a) had the same coupling length as real couplers consisting of \SI{50}{nm} angled filaments. 
A complete ODC device can therefore be integrated into an unprecedentedly small area\cite{Alferness1986,Trinh,Boltasseva2005}, while maintaining reasonably low loss (Fig.~\ref{fig:fig-5-wg}a, blue), therefore achieving a single coupling length system transmission of $0.93$, respectively a loss of \SI{-0.31}{dB}.

% section optical_directional_couplers (end)

%\section{Spectral switching} % (fold)
\label{sec:spectral_switching}

The dispersion of directional couplers is utilized in many compact wavelength division multiplexing (WDM) systems\cite{Yariv1973} as it allows for low-loss wavelength division. Probing the spectral transmission of the investigated ODC circuits reveals a pronounced wavelength-dependence.
We characterized different ODCs spectrally in a range of $\lambda_0 = 1250 - \SI{1850}{nm}$. Starting with $\lambda_0 = \SI{1550}{nm}$, the ODCs with a filament width of \SI{70}{nm} and length of $L = \SI{10}{\micro m}$ (Fig.~\ref{fig:fig-7-odc}b) operating with several full coupling lengths demonstrated a well-balanced output with $I_\text{cross}/I_\text{bar} ≈ 1$ (Fig.~\ref{fig:fig-7-odc}e).

Then, by sweeping the wavelength, we observed full switching between cross (Fig.~\ref{fig:fig-7-odc}d) and bar (Fig.~\ref{fig:fig-7-odc}f) in going from short to long wavelengths, leading to a \SI{30}{dB} wavelength discrimination between \SI{1450}{nm} and \SI{1650}{nm} (see supplementary information for a video demonstrating spectral switching). 

The whole device features a footprint of only $ 5 \times \SI{10}{\micro m^2}$ and could easily be stacked to form an arrayed WDM, enabling even more narrow wavelength discrimination within little more space. The size is extremely compact compared to current dielectric integrated WDMs requiring millimeter dimensions\cite{Asghari2011}.

The strong dispersion of the coupling is dominated by the frequency dependent dielectric constant of the Au as the field-penetration into the thin metallic filament between the waveguide is deep. 

Thus, inside the ODC the total metal-field overlap is large compared to the single waveguide where the metal-dispersion does not influence the effective index of the mode considerably (Fig.~\ref{fig:fig-5-wg}a,~green~curve). 
However, the low dispersion is a desired property for the waveguides as is the high dispersion of the couplers for reducing the effective size of the ODC.

% section spectral_switching (end)

%\section{Conclusions and outlook} % (fold)
\label{sec:conclusions_and_outlook}
In conclusion, we have demonstrated a highly reproducible design scheme for plasmonic nano-circuitry that combines the high confinement and dispersive properties of plasmonics\cite{Boltasseva2005} together with the low loss and high coupling efficiencies of Yagi-Uda antennas. 
For the first time, loaded optical Yagi antennas are used to reduce insertion loss and couple light with a high total efficiency of \SI{60}{\%} (\SI{15}{\%} from air, \SI{45}{\%} from substrate) into and out of plasmonic nano-circuits. 

Furthermore, this platform is used to demonstrate the operation of embedded SPP based ODCs with extraordinarily short coupling length. 
The ODCs feature low transmission loss and compete well with other, e.g. Si integrated circuitry components\cite{Asghari2011,Trinh} in terms of coupling-length-over-loss ratio while being superior in overall compactness.  

Distinct spectral switching is observed, thus allowing down-scaling of wavelength division multiplexing from the millimeter to the micrometer range. 
Additionally, we note that other nano-plasmonic circuitry components can easily be transferred to the Yagi-Uda loaded embedded plasmonic waveguide platform, leading a path towards highly integrated, spectrally functional plasmonic chips like resonant guided wave networks\cite{Feigenbaum2010a} or on-chip detectors\cite{Ly-Gagnon2012}, ideally with no need for subsequent intersections\cite{Delacour2010,Gosciniak2010,Briggs2010} between different types of waveguides.

% section conclusions_and_outlook (end)

%\section{Methods} % (fold)
%\label{sec:methods}
%
%Fabrication
%Measurement
%Evaluation 
%FDTD

% section methods (end)

%% classes.  Rather than use \section, an appropriate macro is
%% provided that will always work.
%%%%%%%%%%%%%%%%%%%%%%%%%%%%%%%%%%%%%%%%%%%%%%%%%%%%%%%%%%%%%%%%%%%%%
\acknowledgement

The authors thank P. Banzer, T. Bauer, S. Dobmann, J. S. Fakonas and H. W. Lee for inspiring discussions and help with a new optical setup. 

This work was supported by the Cluster of Excellence Engineering of Advanced Materials (EAM), Erlangen and the Multidisciplinary University Research Initiative grant (Air Force Office of Scientific Research, FA9550-10-1-0264). A.K. and D.P. also acknowledge funding from the Erlangen Graduate School in Advanced Optical Technologies (SAOT) by the German Research Foundation (DFG) in the framework of the German excellence initiative, A.K. by Friedrich Naumann Foundation and S.P.B. by the National Science Foundation. We acknowledge use of facilities of the Kavli Nanoscience Institute (KNI) at Caltech and the Max Planck Institute for the Science of Light (MPL), Erlangen.  

A.K. and S.P.B. conceived the experiments and developed the device design, A.K. performed numerical simulations and S.P.B. fabricated the samples. A.K., D.P., H.P. and S.P.B. performed the optical and FIB/SEM measurements. A.K., S.P.B., U.P. and H.A.A. analyzed the data and wrote the first draft of the manuscript. All authors contributed to the final version of the manuscript.

%% \suppinfo macro.
%%%%%%%%%%%%%%%%%%%%%%%%%%%%%%%%%%%%%%%%%%%%%%%%%%%%%%%%%%%%%%%%%%%%%
\suppinfo

Experimental details on the fabrication the farfield optical IR setup and characterization procedures, leakage microscopy measurements, spectral switching, NSOM measurements and experimental data analysis and statistics on the antennas and directional couplers and numeric details on the FDTD simulations.

\bibliography{own-paper_embedded_components_ln_mendeley}

\end{document}

%%%%%%%%%%%%%%%%%%%%%%%%%%%%%%%%%%%%%%%%%%%%%%%%%%%%%%%%%%%%%%%%%%%%%
%% The manuscript does not need to include \maketitle, which is
%% executed automatically.  The document should begin with an
%% abstract, if appropriate.  If one is given and should not be, the
%% contents will be gobbled.
%%%%%%%%%%%%%%%%%%%%%%%%%%%%%%%%%%%%%%%%%%%%%%%%%%%%%%%%%%%%%%%%%%%%%


\section{Optical setup for far-field characterization } % (fold)
\label{sec:setup}

\begin{figure}
	  \includegraphics{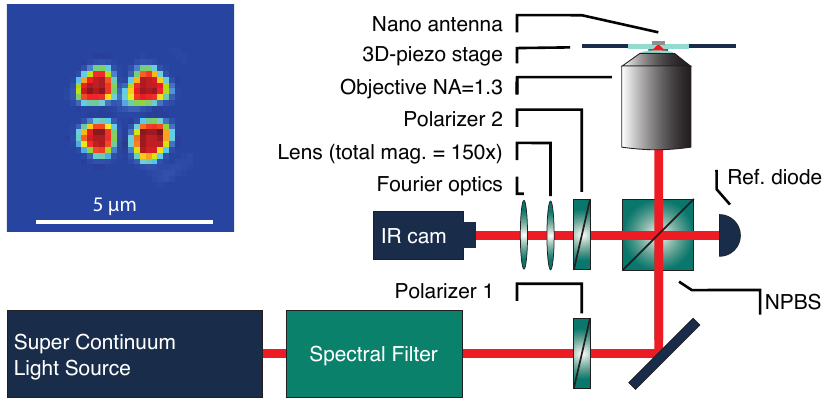}
	  \caption{The experimental setup to probe the polarization dependent spectral emission of a nano-plasmonic circuit in real and Fourier space for a broad spectral range $\lambda = \SI{1200}{nm}$ to \SI{1850}{nm}. Inset: The emission from the antenna through crossed polarizers.}
	  \label{fig:setup}
	\end{figure}

	The nano-circuits were positioned with 5 nm xyz precision with a piezo scanning system (Physik Instrumente, PI) and optically characterized with a custom-made optical setup (\ref{fig:setup}) that follows the design published by Banzer et al \cite{Banzer2010a}. 
	
The collimated beam from a supercontinuum light source is spectrally filtered by a programmed acousto-optic tunable filter (operated at $\lambda = 1200 - \SI{1850}{nm}$) (NKT Koheras, SuperK Extreme) and is subsequently directed through a polarization filter (1) and a non-polarizing beam splitter (NPBS), that directs \SI{50}{\%} of the power to a reference diode (InGaAs). The main beam is focused with a high NA objective ($\text{NA} = 0.9$ from air and $\text{NA} = 1.3$ immersion, from substrate) that was carefully characterized to preserve the polarization properties of the laser beam. The diameter of the collimated beam was carefully measured ($FWHM = \SI{1.6}{mm}$) with an InGaAs camera and the effective numeric aperture of the experimental focal spot was determined for all subsequent evaluation steps. The focus of the objective is adjusted on the investigated excitation nano-antenna.

At the same time, the objective images the complete nano-circuit, including the emission from the other antennas. The collimated beam, carrying the image, is polarization filtered (polarizer 2) and passes imaging optics to form a real image on an InGaAs NIR CCD camera (Xenics XS, 320 x 256 pixels). The setup features a variable magnification factor (150x, 300x) and an additional, switchable focusing unit to allow for Fourier plane imaging.

By adjusting the polarization filter (2) perpendicular to polarization filter (1), the reflection of the incident beam is mostly cancelled in the image ($1:5000 - 1:10000$ extinction ratio) and the emission from the monitor antennas of the circuit, which are \ang{90} turned with respect to the excitation antenna, can be measured with exceptionally high aspect ratio and dynamic range. However, conversion of field components in the highly focused beam into the perpendicular polarization causes a characteristic four-lobe-pattern that remains intact even when the incident beam is focused on an antenna with no fabrication errors (\ref{fig:setup} inset). This four-lobe pattern is visible in all measurements at the position of the incoupling antenna (Fig. 7d,e,f of the letter).

The absolute transmission of the nano-circuits was determined by normalization with the known reflectivity of the sample substrate. The spot intensity of the monitor antennas (Fig. 7b of the letter, green circles) was integrated. For each measurement picture for a certain wavelength and circuit, a reference image was taken, featuring the pure reflection of the four-lobe pattern on the substrate. Quasi-analytically, the reflectivity of a silica-to-air interface was calculated by developing a focused Gaussian beam with the correct NA into plane waves, taking into account the crossed second polarization filter. 

Possible fluctuations of the probe laser between the main and the reference measurement are eliminated with the reference diode (\ref{fig:setup}). This method of absolute normalization eliminates sources for additional measurement errors and cancels dispersion and loss of the optical setup.

\begin{figure}
  \includegraphics{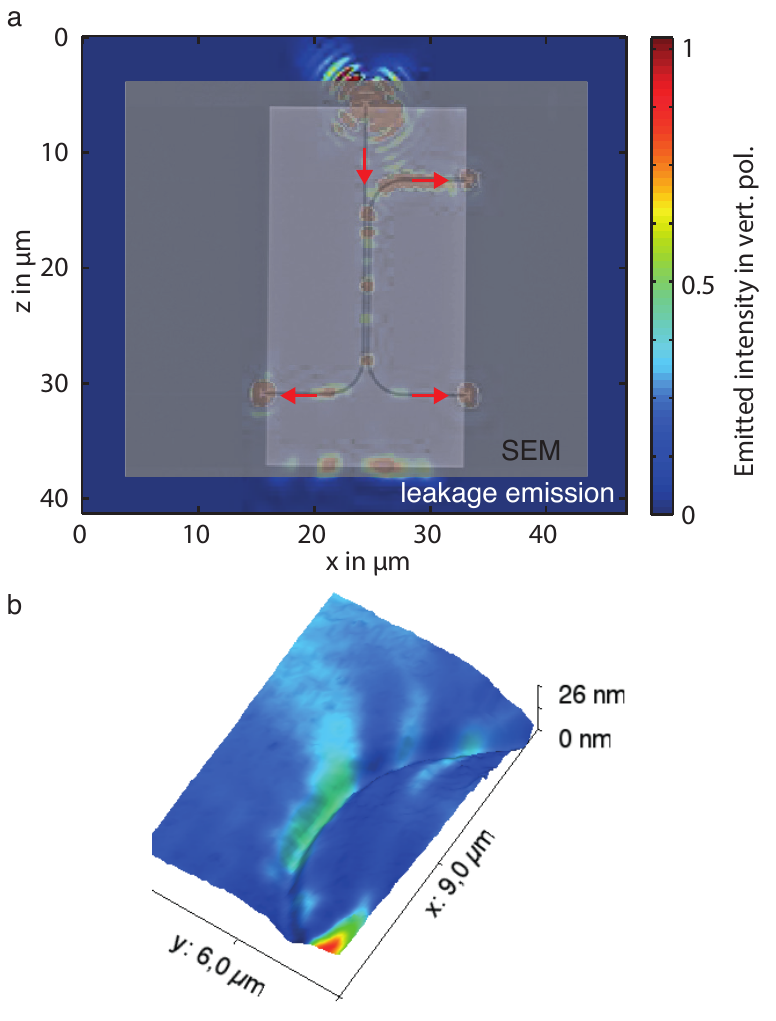}
   \caption{Leakage microscopy of the emission from a nanocircuit with a directional coupler in the configuration with crossed polarizers ($5000  \leq \text{suppression} \leq 10000$). Highly ($> 300 \times$ saturation intensity) overexposed, this real-plane image shows strongly saturated emission from the incident (top) and monitor (left, right) antennas.}
  \label{fig:sem-leakage}
\end{figure}

% section section_name (end) 

\section{Leakage microscopy} % (fold)
\label{sec:leakage_microscopy}

The guided mode of the waveguides does intrinsically not emit into the far-field. However, every waveguide has slight imperfections. Utilizing the high dynamic range of the optial far-field setup, it is possible to image the plasmonic nano-circuit with leakage-microscopy. Highly overexposed ($> 300 \times $saturation intensity) images of a circuit, featuring a directional coupler (ODC) in a direct overlay with an SEM of the same structure show beating patterns inside the waveguides and the coupler.

However, for this high exposure, even the reference antenna (compare to Fig. 7b, blue) shows some emission. The slight scattering at the lower edge of the circuit metal pad indicates that slight residual bend loss is due to the conversion into planar SPPs. These subsequently propagate along the metal surface, are scattered out at the lower edge and detected with very high efficiency due to their polarization being parallel to the emission polarization of the monitor antennas.

% section leakage_microscopy (end)

\section{Near Field Optical Scanning Microscope} % (fold)
\label{sec:near_field_optical_scanning_microscope}
The near field scanning optical microscopy (NSOM) measurements (Fig. 4b) were taken with a custom made setup for polarization-adjusted coupling into the optical antennas, based on a modified commercial fiber aperture NSOM system (Nanonics MV 4000). 

With tapping, phase-controlled AFM feedback, a   $ \text{\O} \approx \SI{270}{nm}$ Au/Cr coated, FIB processed fiber aperture tip was raster-scanned across the area around the antenna ($10 \times \SI{10}{\micro m}$) with 40 nm lateral scan resolution. The near-field, collected from the structure and converted into a propagating mode of an optical fiber, was detected with a high sensitivity, highly amplified InGaAs diode detector. 

To reveal the guided mode inside the connected waveguide the scanned intensity is displayed with log-scale and saturation at the beam position was accepted. The plane of scanning is leveled 320 nm above the upper metal boundary due to the cladding spin-on glass (SOG) layer, which in the topography channel of the NSOM still showed smoothed variations of height within $\Delta z \approx \SI{20}{nm}$ that are caused by the underlying $ \Delta z \approx \SI{220}{nm}$ topography of the metal structures. 
In this measurement already little maladjustment of the laser focus leads to slight diffraction artifacts towards the metal surface. To ensure precise adjustment, during the scan also the reflected light from the antenna was collected through the excitation optical pathway and monitored to quantify the distortion of the reflected signal by the scanning NSOM tip.

Additional measurements showed low additional loss from waveguide bends with at least $R = \SI{3}{\micro m}$ bend radius (Fig. S2b), which coincides well with the far-field circuit characterization that indicated these bends to be applicable with negligible additional loss compared to the straight waveguide propagation loss.

% section near_field_optical_scanning_microscope (end)

\section{Fitting the circuit transmission with a lossy Fabry-P\'{e}rot model} % (fold)
\label{sec:fitting_the_circuit_transmission_with_a_lossy_fabry_perot_model}

Based on the lumped network circuit model\cite{Engheta2007} (Fig. 7a) the nano-circuit was treated as a Fabry-P\'erot resonator to determine waveguide and antenna properties in one iterative least-square fit over the independent variables wavelength and waveguide length.
The fitted model is based on the transmission of a Fabry-P\'erot resonator with loss in the resonator and at the mirrors, which has length $L$ and a characteristic wave number $k$

$$ T_\text{FPR} =  \frac{T_\text{ant}  }{ 1 + \Gamma \sin^2 (kL) },$$ 

where $T_{\text{ant}}$ represents the efficiency of the antennas to couple the focused laser beam into the FPR cavity, respectively out of it. The antennas are assumed to experience Lorentzian broadening due to plasmonic loss:

$$T_\text{ant} = \frac{ T_\text{ant}^\text{max} } {1+ \left( \frac{\lambda-\lambda_0}{\gamma} \right)^2 }  .$$ 

$\Gamma$ determines the inverse line width that can be expressed in terms of the reflectivity $R = R_\text{in} = R_\text{out}$ of the resonator mirrors, which is caused by imperfect impedance matching between the optical antennas and the waveguides

$$ \Gamma = \frac{ 4 \,R \, e^{-\alpha L} }{ \left( 1 - R \, e^{-\alpha L} \right)^2}   . $$

The antenna efficiency curve and the antenna loss curve are caused by the same loss mechanism. Hence, the reflectivity of the antenna resonator mirrors can be described as
\begin{align*}
R   & = 1-T_\text{ant}-A_\text{ant}  \\
    & = 1- \frac{ T_\text{ant}^\text{max}+A_\text{ant}^\text{max} } { 1+ \left( \frac{\lambda-\lambda_0 }{\gamma}  \right)^2 }
\end{align*}

where, according to the applied model, $A_\text{ant}$ represents the absorption inside the antenna with 
$A = 1 - T - R$. 
With an iterative fit of this model to the experimental circuit transmission values, the spectral propagation length $L_0$, the effective index of the guided mode ( $n_\text{eff} = k \lambda_0/(2\pi) $ ), the antenna efficiency $T_\text{ant}^\text{max}$, the central wavelength $\lambda_0$ and antenna linewidth $\text{FWHM} = 2 \gamma$ are determined as free parameters. There may be additional sources for loss, which we are not able to systematically determine. 

We can however determine an upper bound for the loss $A \leq 1 - T_\text{min} - R$, which effectively makes $T_\text{ant}(\lambda)$ a lower bound for the spectral transmission efficiency of the antennas to convert light from the far field into the waveguide and vice versa.
Error propagation was applied to all determined free variables $T_\text{ant}^\text{max}, \gamma, \lambda_0, L_0, n_\text{eff}$, based on the error bounds from the iterative fit over several measurements and different fabricated structures. The determined values for the standard deviation $\sigma$, as indicated in Fig. 2b, 5a and 5b therefore include statistical errors of fabrication and measurement.

% section fitting_the_circuit_transmission_with_a_lossy_fabry_perot_model (end)

\section{Fitting the optical directional coupler transmission model} % (fold)
\label{sec:odc}

The directional couplers are well described with coupled mode theory. The coupling length $L_\text{c}$ for periodic full transfer of the power from one waveguide to the other was determined by fitting this model to experimental data over a variation of the parallel length of the couplers $L = 0, 3, 6, 9,\SI{12}{\micro m}$. 
The system is well described as a weakly coupled system of two harmonic oscillators, exchanging energy proportional to the coupling constant $\kappa$. The intensity out of the waveguides bar and cross after an ODC of length $L$ is given by
$$I_\text{bar} = \cos^2 (\kappa L) \text{ and }  I_\text{cross} = \sin^2 (\kappa L)$$

\begin{figure}
  \includegraphics{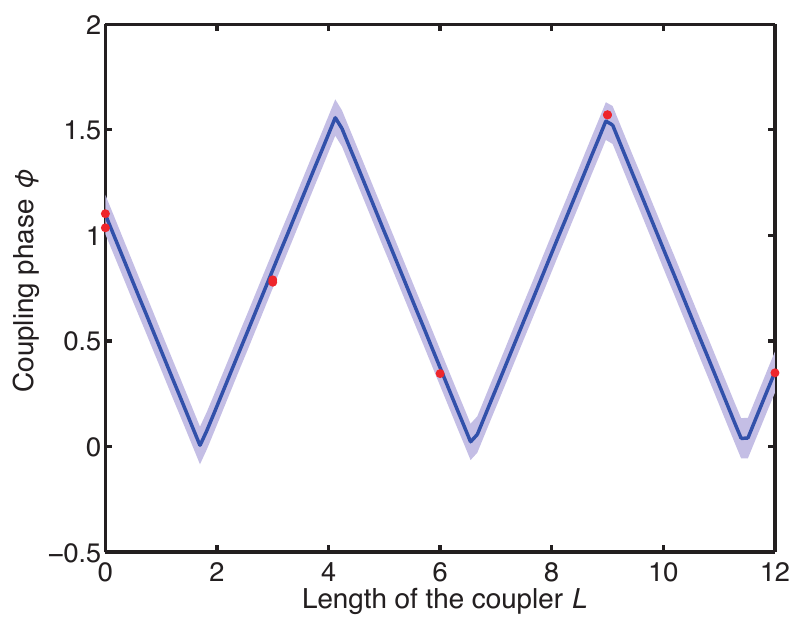}
  \caption{The experimental coupling ratio of 8 nano-circuits with directional couplers of varied straight length $L$ (all $w = \SI{50 }{nm}$) is converted to the periodic coupling phase (red dots) and is iteratively fitted with coupled mode theory. Fit result (blue curve) and 90\% confidence intervals (blue shaded).}
  \label{fig:odc-fit}
\end{figure}

and the coupling length can be expressed as $L_\text{c} = \pi / 2 \kappa$. We used the ratio of the integrated emission from both spots of the monitor antennas (Fig. 7b green, compare Fig. 7 d,e,f)) $I_\text{cross}/I_\text{bar}$ as it is intrinsically more robust to measurement error and even slight maladjustments of the exciting laser beam. However, $I_\text{cross}/I_\text{bar}=\tan^2 (\kappa L)$ diverges and is unsuitable for iterative least-square fitting. A more suitable approach was chosen by fitting the non-diverging coupling phase, which is as well periodic with full coupling lengths:
\begin{align*}
\varphi(L) &= \arctan⁡ \left( \frac{I_\text{cross} }{I_\text{bar} }  \right)\\
		   &= \left| \left[ \pi \frac{ L-L_\text{bend} }{2 L_\text{c} }\right]_\text{mod}^\pi - \frac{\pi}{2} \right|  
\end{align*}

This expression allows for a fit over a variation of the length of the coupler $L$. $L_\text{bend}$ represents an additional offset, which is caused by coupling that already occurs in the bends where cross and bar waveguides approach with a radius of $R = \SI{3}{\micro m}$. This bend-coupling is expressed in terms of an extra virtual length of the coupler and leads to significant transfer of energy even in the case of zero straight parallel length of the ODC ($L = 0$).

The investigated circuits with the lengths available already cycle through several full coupling lengths. The ODCs with the most narrow filament ($w = \SI{50}{nm}$) resulted in the best fits with least statistical fabrication and measurement deviations and the shortest coupling length, as displayed in Fig. S3. In this case eight measurements on different fabricated circuits were combined. The resulting fit is displayed with 90\% prediction bands and gave $L_\text{c} = 2.46 \pm \SI{0.04}{\micro m}$ coupling length and $L_\text{bend} = 1.7 \pm \SI{0.07}{ \micro m}$ equivalent coupler length of the bend. The indicated error margins take account for statistical errors, not potential systematic errors that might not have been possible to determine with the applied methods.

% section odc (end)

\section{Simulation with FDTD and particle swarm optimization (PSO)} % (fold)
\label{sec:simulation_with_fdtd_and_particle_swarm_optimization_pso_}

The antennas, waveguides and couplers were simulated with a 2D and full 3D Finite Difference Time Domain solver (Lumerical FDTD Solutions). For design purposes, literature values from Johnson and Christy\cite{Johnson1972} were used as material parameters. After first fabricated structures were experimentally characterized, the simulations were repeated with the ellipsometrically determined dielectric susceptibility of Au, which turned out to be close to the initially used literature values in the real part of the dielectric susceptibility, but deviated up to a factor of $1.5$ in the imaginary part.

In the simulations, the antenna was excited with a highly focused Gaussian beam with the same beam properties as determined for the experiment with the objectives in use (NA = 0.9 from air and $\text{NA} = 1.3$ from silica). Its emission directionality and efficiency were analyzed in reverse configuration, exciting the guided mode inside the SPP gap waveguide with a mode solver and observing the emission with appropriate frequency domain ports. The more narrow resulting angular directivity of the such optimized antennas was experimentally taken into account by adjusting the collimated beam diameter to slightly reduce the numeric aperture and was characterized as described in section~1.

The supplementary video shows the electric field $|E|^2$ in time domain, as a focused beam ($\text{NA} = 0.9$) impinges the optimized geometry of a Yagi-Uda nano-antenna and is converted into a guided mode, propagating along the waveguide.

% section simulation_with_fdtd_and_particle_swarm_optimization_pso_ (end)

\section{Instrumentation} % (fold)
\label{sec:instrumentation}

All investigated nano-circuits were fabricated with a lift-off patterning process as described in the main article. A 100 kV Leica EBPG-5000+ system with pattern generator transferred the nano-circuit designs to the resist. Standard procedures for development, lift-off and standard manufacturer-recipes were used for the application, baking and curing under $\text{N}_2$ atmosphere of the commercially available silsesquioxane spin-on-glass (SOG) Filmtronics 400F.
For cross-sections, geometric characterization and fabrication optimization a Zeiss Dual Beam FIB system and a FEI Nova-200 FIB system were used.
The laser for far-field and near-field measurements was a NKT Photonics / Koheras SuperK Extreme supercontinuum source with nominal \SI{6}{W} output power and subsequent NKT Koheras acousto-optical tunable filters (AOTFs) for tunable wavelength-selection.
A sensitive InGaAs NIR camera (Xenics CS, $320 \times 256$) pixels was used for image acquisition of the light emitted from the circuits.

% section instrumentation (end)

\bibliography{supplementary-ln}